\documentclass[10pt]{iopart}
\usepackage{graphicx}
\usepackage{xspace}				
\usepackage[pdftex]{color}			

\newcommand{\bma}[1]{\mbox{\boldmath$#1$}}
\newcommand{\Cdot}{\bma{\cdot}}
\newcommand{\Nabla}{\bma{\nabla}}

\newcommand{\lisa}{\textsl{LISA}\xspace}

\newcommand{\ltp}{\textsl{LTP}\xspace}

\begin{document}


\jl{6}

\title[The LISA PathFinder DMU and Radiation Monitor]{The LISA
PathFinder DMU and Radiation Monitor}

\author{P~Canizares$^{1,4}$, A~Conchillo$^{1,4}$, M~D\'\i az--Aguil\'o$^{2,4}$,
 E~Garc\'\i a-Berro$^{2,4}$, L~Gesa$^{1,4}$, F~Gibert$^4$,C~Grimani$^3$,
 I~Lloro$^{1,4}$, A~Lobo$^{1,4}$, I~Mateos$^{1,4}$, M~Nofrarias$^5$,
 J~Ramos-Castro$^6$, J~Sanju\'an$^7$ and Carlos~F~Sopuerta$^{1,4}$}

\address{$^1$ Institut de Ci\`encies de l'Espai, {\sl CSIC}, Facultat de
 Ci\`encies, Torre C5 parell, 08193 Bellaterra, Spain}

\address{$^2$ Departament de F\'\i sica Aplicada, {\sl UPC},
 c/ Esteve Terrades 5, 08860 Castelldefels, Spain}

\address{$^3$ Universit\`a degli Studi di Urbino, and {\sl INFN\/} Florence,
 Istituto di Fisica, Via Santa Chiara 27, 61029 Urbino, Italy}

\address{$^4$ Institut d'Estudis Espacials de Catalunya (IEEC), Edifici Nexus,
 Gran Capit\`a 2-4, 08034 Barcelona, Spain}

\address{$^5$Max-Planck-Institut f\"ur Gravitationsphysik
 (Albert-Einstein-Institut), Callinstrasse 38, D-30167 Hannover, Germany}

\address{$^6$Departament d'Enginyeria Electr\`onica, UPC, Campus Nord,
 Edifici C4, Jordi Girona 1-3, 08034 Barcelona, Spain}

\address{$^7$Department of Physics, University of Florida, NPB--22258
 PO~Box~118\,440, Gainesville FL 32611-8440, USA}

\ead{lobo@ieec.fcr.es}

\date{11-August-2008}

\begin{abstract}
The LISA PathFinder DMU (Data Management Unit) flight model was
formally accepted by ESA and ASD on 11 February 2010, after all
hardware and software tests had been successfully completed. The
diagnostics items are scheduled to be delivered by the end of 2010.
In this paper we review the requirements and performance of this
instrumentation, specially focusing on the Radiation Monitor and the
DMU, as well as the status of their programmed use during mission
operations, on which work is ongoing at the time of writing.
\end{abstract}
\noindent\emph{Keywords}: \lisa, \lisa Pathfinder, gravitational wave
detector, charge deposition, Radiation Monitor, Data Management Unit
\pacs{04.80.Nn, 95.55.Ym, 04.30.Nk,07.87.+v,07.60.Ly,42.60.Mi}
\submitto{\CQG}
%

\section{Introduction}
\label{lobo-sec1}

LISA is a technologically sophisticated mission. So complex, in fact,
that ESA decided to fly a precursor mission to ensure technology
readiness and maturity have reached a safe status to start LISA.
Launch for PathFinder is currently set for early 2013.

%

LISA PathFinder (LPF) has a reduced acceleration noise budget, both in
magnitude and in frequency band with respect to LISA~\cite{lobo-lpfsrd}:
\begin{equation}
 S_{\delta a, {\rm LPF}}^{1/2}(\omega)\leq 3\!\times\!10^{-14}\,\left[
 1 + \left(\frac{\omega/2\pi}{3\ {\rm mHz}}\right)^{\!\!2}\right]\,
 {\rm m}{\rm s}^{-2}/\sqrt{\rm Hz}
 \label{lobo-eq2}
\end{equation}
in the frequency band between 1 mHz and 30 mHz.

This noise can be apportioned to different components, each of them
individually interfering with the mission performance. Requirements
are set on each of those so that the total sum is compatible with
equation~(\ref{lobo-eq2})~\cite{lobo-lpfsrd}. Thermal, magnetic and
charged particle flux are specifically monitored by dedicated hardware
and software, forming what is known as the \emph{Data and Diagnostics
Subsystem} (DDS), which also includes the DMU (\emph{Data Management Unit}),
the LTP (LISA Technology Package, or mission payload) computer. The
DDS has been designed and built in IEEC-CSIC in Barcelona, Spain,
and Flight Models of each of its parts have been delivered, or
will be by the end of 2010, after passing extensive tests to ensure
their performance during mission operations.

In this paper we review the latest progress with the DDS parts,
specially focusing on the Radiation Monitor and the DMU, which have
been the subject of our work during the last few months. We will
however give a quick summary of the other diagnostics items as
well in the first sections; a more detailed description of them
can be found in reference~\cite{lobo-lisa7}.

\section{Thermal diagnostics}
\label{lobo-sec2}

The thermal stability in the LTP Core Assembly (LCA) has to meet a
very stringent requirement:
\begin{equation}
 S_{\delta T}^{1/2}({\omega}) \leq 10^{-4}\,{\rm K\,Hz}^{-1/2}\ ,\quad
 1\,{\rm mHz}\,\leq\,\frac{\omega}{2\pi}\,\leq\,30\,{\rm mHz}
 \label{lobo-eq3}
\end{equation}

A set of 24 temperature sensors are scattered around the LCA which
are intended to measure temperatures at as many strategic spots.
More specifically, in the outer walls of the Electrode Housings of
both Test Masses (TM), at both Optical Windows (OW), at the Optical
Bench (OB) and in the suspension struts (SS). In order to be able
to make significant measurements in an environment with such a
degree of thermal stability as required by equation~(\ref{lobo-eq3}),
an even more demanding requirement must be set on the performance
of these sensors. This is~\cite{lobo-fee}
\begin{equation}
 S_{\delta T {\rm ,\ sensors}}^{1/2}({\omega}) \leq 10^{-5}\,{\rm K\,Hz}^{-1/2}\ ,
 \quad 1\,{\rm mHz}\,\leq\,\frac{\omega}{2\pi}\,\leq\,30\,{\rm mHz}
 \label{lobo-eq4}
\end{equation}

\begin{figure}
\centering
\includegraphics[width=0.8\columnwidth]{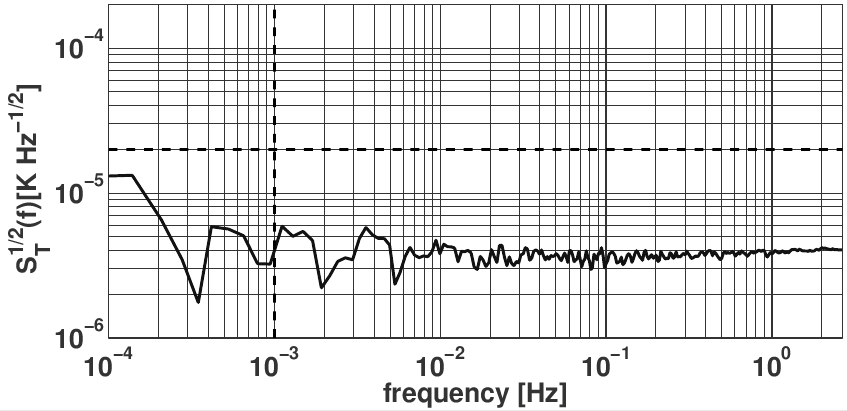}
\caption{Flight model temperature sensor behaviour. The plot corresponds
to a differential measurement between two NTCs. As can be seen, it is
fully compliant with the requirement in equation~(\protect{\ref{lobo-eq4}}).
\label{lobo-fig1}}
\end{figure}

The temperature sensing in the DDS is done by means of NTC (Negative
Temperature Coefficient) \emph{thermistors}, because they have more
pronounced slopes in the temperature--resistance plane than other
devices such as Platinum resistors. In addition, an extremely quiet
front-end electronics must be used to drive them and acquire data.
This electronics is part of the DMU, and Figure~\ref{lobo-fig1}
shows the performance of a differential temperature measurement
between two NTCs as recorded in the latest tests done with Flight
Model sensors and DMU. This was April-2010. A detail must however
be clarified: because of mass and power budget limitations, not
every one of the 24 thermistors can have its own electronic board;
rather, they are \emph{multiplexed} in six groups of four. While
this makes the system globally compliant with those constraints,
it introduces additional noise, as the real sampling frequency is
divided by 4 in each thermistor. Therefore the multiplexed system
noise is higher by a factor of 4$^{1/2}$\,=\,2 than shown in the
plot. Note however that we are still fully within requirements.
Actually, the (individual) NTC noise is below 10$^{-5}$\,K\,Hz$^{-1/2}$
all the way down to 10$^{-5}$\,Hz, i.e., well within the LISA band.
This means the current LTP thermal system can probably be transferred
basically as is into LISA.

\subsection{Precision heaters}
\label{lobo-sec2.1}

It needs to be stressed that such marvellously working temperature
sensors are not by themselves of much use for the purpose they will
be embarked. As explained in the previous section, we need these
monitors to record temperature fluctuations which affect the
performance of the LTP. But how do we link a (known) temperature
fluctuation with an (unknown) TM shake or an interferometer phase
jitter? Although this link can be modelled, it has to also be established
in flight by direct measurement ---since the model may be inaccurate.

To the effect, a set of 14 \emph{precision heaters} are also installed
in the LTP at some suitable spots~\cite{lobo-heaters}. Their role is
to inject controlled thermal signals in the system such that they are
strong enough to be clearly seen in the LTP readout, i.e., with a
signal-to-noise ratio around 100. Temperature measurements are
taken as well and hence a relationship is established between the
latter and the observed response of the interferometer and/or the
Gravitational Reference Sensor (GRS). The analysis is done in frequency
domain, so that a set of transfer functions is obtained relating
temperature variations to LTP response. Finally, the hypothesis
is made that the behaviour of the LTP is maintained when the
heaters are switched off.

Heaters are driven with very low noise DMU electronics. There are
two major groups of heaters: those in the GRS and those elsewhere
\cite{lobo-heaters}. The ones in the second group are rather
conventional \emph{kapton} heaters, consisting in a wire winding
held together by an elastic resin. These heaters, however, cannot
be attached to the walls of the Electrode Housing (EH) of the GRS
due to their magnetic properties: they actually have some significant
quantities of ferromagnetic elements, notably Nickel, which are likely
to compromise the stability of the TM. It was therefore decided that
NTCs should \emph{also} be used as heaters in that part of the LTP.

These NTCs are a bit more difficult to drive than the others due
to their changing electric resistance upon being heated up. Indeed,
when a voltage is applied to an NTC the current through it will heat
it up, hence its resistance will drop, hence the current and power it
dissipates will increase. It has been shown~\cite{lobo-pephd} that,
under suitable conditions of initial NTC temperature, thermal
resistance between the heater and the attachment block, and
voltage range, this process ends up in a stationary state in
a matter of about 10~seconds. This is good enough for the time
scales of the heating operations (normally a few thousand seconds),
but the DMU has to be programmed such that these heaters be
commanded to deliver the required power in the stationary state.
In other words, the voltage may not be just $V^2/R$, where $R\/$
is some nominal NTC resistance value, but rather $V^2/R_\infty$,
where $R_\infty$ is the NTC stationary state resistance, instead.
Actually the relationship between voltage applied and stationary
state power dissipation is somewhat involved:
\begin{equation}
 P_\infty = \frac{V^2}{R(T_\infty)} =
 \frac{V^2}{R\left(T_0+\frac{\displaystyle\theta V^2}%
 {\displaystyle R_0+\alpha_0\theta V^2}\right)}
 \label{lobo-eq5}
\end{equation}
where $T_0$ is the NTC temperature just prior to activation (measured
with an adjacent sensor), $\theta\/$ is the above mentioned thermal
resistance, and $\alpha$\,=\,$R^{-1}\,dR/dT\/$ is the thermistor's
temperature coefficient. The dependence of $T\/$ on $R\/$ is
usually given by the Steinhart-Hart equation~\cite{lobo-sh},
again a somewhat involved mathematical expression.

The algebraic complexities of the above formulae make it impossible
to invert equation~(\ref{lobo-eq5}) to obtain the voltage $V\/$ as
a function of the power $P_\infty$ in a closed-form expression. This
circumstance has motivated the use of Look-Up Tables (LUT) in the DMU
software which commands the GRS heaters: for a given thermal resistance
$\theta\/$ (measured on ground) and an NTC temperature (measured
dynamically on flight), the SW looks for the tabulated power which
is closest to the one required for injection, then reads off the
LUT the activation voltage, and applies it~\cite{lobo-aleix}.

\section{Magnetic diagnostics}
\label{lobo-sec3}

The LTP Test Masses are made of a 70\,\% Au + 30\,\%\ Pt alloy, a good
combination to keep as low as possible their magnetic properties and,
at the same time, to provide sufficient mechanical robustness to
withstand launch shaking while caged. It is however impossible to
avoid the presence of ferromagnetic residuals after the casting,
hence some magnetic remanence will be there. Limits have been set
on remnant magnetic moment ${\bf m}_0$ and susceptibility $\chi$
as~\cite{lobo-lisa7}
\begin{equation}
 |\chi| < 10^{-5}\ ,\quad
 |{\bf m}_0| < 10^{-8}\ {\rm Am}^2
 \label{lobo-eq6}
\end{equation}

These parameters couple to the surrounding magnetic field, mostly
created by spacecraft electronic boxes and other components, such
as solar panels, FEEP (Field Effect Electric Propulsion, the
satellite's micro-thrusters), etc., thereby creating forces and
torques on the Test Masses, and fluctuations thereof, i.e., magnetic
noise. According to standard electromagnetic theory, forces and
torques are respectively given by
\begin{equation}
 {\bf F} = \left\langle\left[\left({\bf M} +
 \frac{\chi}{\mu_0}\,{\bf B}\right)\Cdot\Nabla\right]%
 {\bf B}\right\rangle V
 \label{lobo-eq7}
\end{equation}
and
\begin{equation}
 \textbf{N} = \left\langle\textbf{M}\times\textbf{B} +
 \textbf{r}\times \left[\left(\textbf{M}\Cdot\Nabla\right)\,\textbf{B} +
 \frac{\chi}{\mu_0}\left(\textbf{B}\Cdot\Nabla\right)\textbf{B}\right]
 \right\rangle V
 \label{lobo-eq8}
\end{equation}
The meaning of symbols in the above formulae is as follows:

\begin{center}
\begin{tabular}{lcl}
 {\bf B} & \ & Magnetic induction field in the TM \\
 {\bf M} & & Density of magnetic moment (\emph{magnetisation}) of the TM \\
 {\bf r} & & Vector distance to the TM centre of mass \\
 $V$ & & Volume of the TM \\
 $\chi$ & & Magnetic \emph{susceptibility} of the TM \\
 $\mu_0$ & & Vacuum magnetic constant (4$\pi$$\times$10$^{-7}$
 m\,kg\,s$^{-2}$\,A$^{-2}$)
\end{tabular}
\end{center}
while $\langle\cdots\rangle$ indicates TM volume average of enclosed
quantity. For example, for any magnitude $f({\bf x})$, such average is
defined by the volume integral
\begin{equation}
 \langle f\rangle\equiv\frac{1}{V}\,\int_V\,
 f({\bf x})\,d^3x
 \label{lobo-eq9}
\end{equation}
over the TM volume. Magnetic noise can be readily inferred from the
above formulae. There is no specific requirement on either magnetic
field or gradient fluctuations, there is only one on the total
contribution of magnetic disturbances to the overall mission
acceleration noise budget; magnetic is required to stay below
40\% of that budget~\cite{lobo-lpfsrd}. On the other hand, there
are specific requirements on the DC values of magnetic field and
gradient, which should be kept below 10\,$\mu$-Tesla and 5\,$\mu$-Tesla/m,
respectively. This is because of the quadratic dependence of the force
{\bf F} on the magnetic field {\bf B}, due to the presence of a
non-zero susceptibility ---see equation~(\ref{lobo-eq8}); when it
comes to evaluating fluctuations, it is immediately seen that DC
field values couple to gradient fluctuations, and also DC gradient
values couple to field fluctuations.

Magnetic fields in the LTP are monitored by means of four tri-axial
fluxgate magnetometers, located in the periphery of the LCA. These
are not ideally suited to infer the magnetic field and gradient at
the TMs, due to their distance to the latter. Classical interpolation
methods have failed to produce acceptable results, but \emph{neural
network} algorithms have been developed and studied at IEEC which
produce significant improvements in our ability to determine field
values at the TMs~\cite{lobo-neural}. The reader is also recommended
to look into Marc D\'\i az's contribution to this Proceedings for
more in-depth analysis and details.

\subsection{Alternative solutions}
\label{lobo-sec3.1}

\begin{figure}
\centering
\includegraphics[width=0.8\columnwidth]{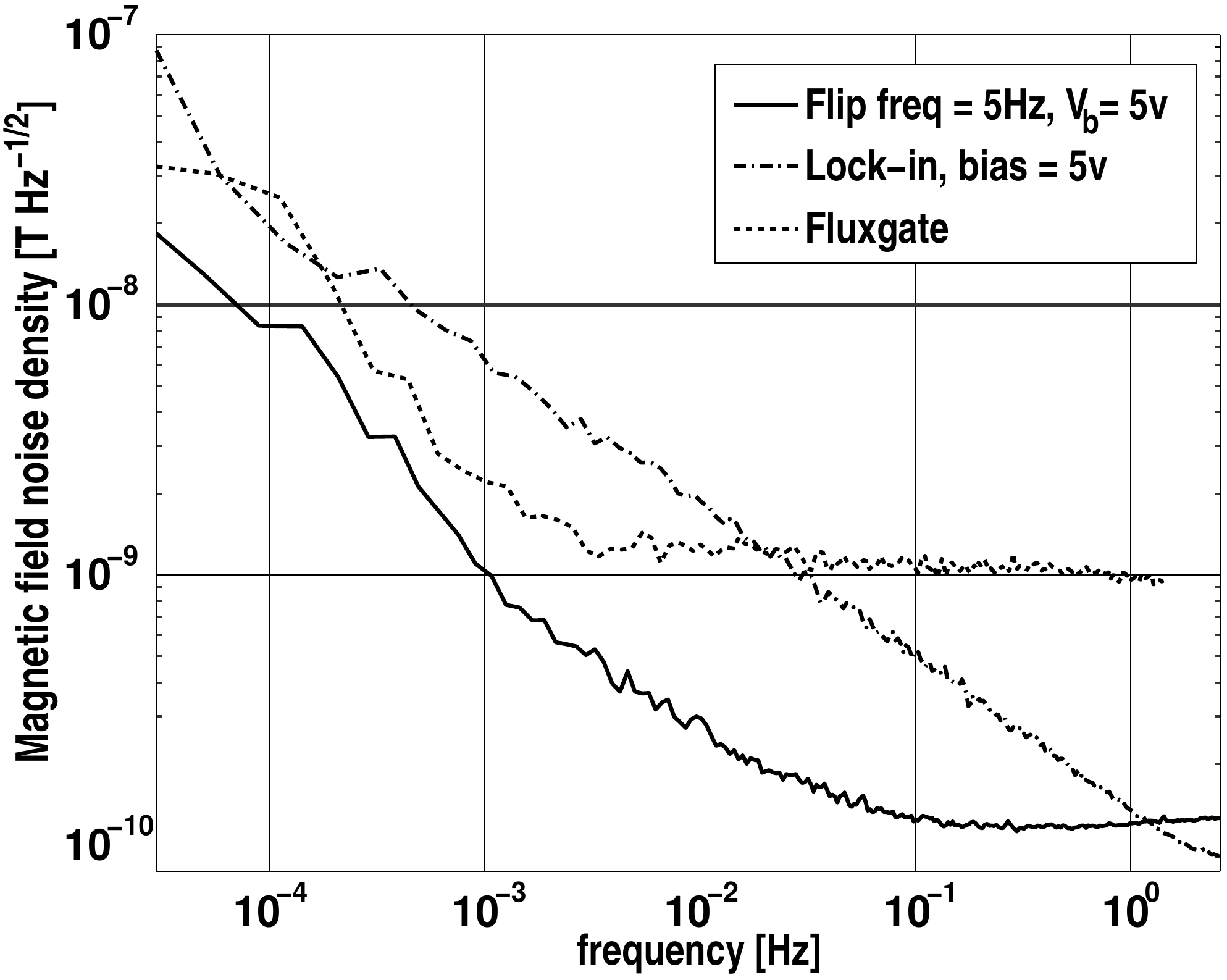}
\caption{Magnetometers sensitivity curves. The horizontal, thicker grey
line at 10$^{-8}$ Tesla\,Hz$^{-1/2}$ is the current LPF magnetometer noise
requirement, which is comfortably met in all cases considered, i.e.,
for frequencies above 1\,mHz. The curves show that the best performance
corresponds to the AMR with a \emph{flip} voltage scheme, showing
significant improvements over the more conservative lock-in approach.
The Fluxgate performs well but is slightly noisier at LISA frequencies.
\label{lobo-fig2}}
\end{figure}

While it is not possible to make changes now to the general LPF plan,
e.g., magnetometers cannot be replaced, it would be unwise to think
of LISA carrying on board a set of a few high performance but distant
and voluminous fluxgate magnetometers. Ongoing research at IEEC has
revealed that alternatives to the LPF scheme seem to exist. These
are based on both a new type of sensors and driving electronics.
The new sensors are tiny AMR (anisotropic magneto-resistor), while
the electronics resorts to periodic flipping voltages to apply
set-reset signals to the devices, thereby enhancing their performance.
In Figure~\ref{lobo-fig2} we see noise curves of various magnetic
sensors and activation techniques which show the significant
advantage which can be potentially drawn from the new AMRs. There are
issues still to be properly addressed, perhaps most notably the possible
magnetic back-action of the sensors, since they have tiny ferromagnetic
cores. Analysis so far point towards a negligible such back-action, the
reader will find much more detailed information in Ignacio Mateos's
contribution to this Conference Proceedings.

\subsection{Control coils}
\label{lobo-sec3.2}

Just like thermal diagnostics require precision heaters, so magnetic
diagnostics require precision induction coils. Their purpose is very
similar to that of the heaters, though with a difference: strong
magnetic signals injected by the induction coils, together with the
LTP response to them (again in the order of SNR\,$\simeq$\,100) can
be used to determine the remnant magnetic moment and the susceptibility
of the TMs. More information can be found in~\cite{lobo-lisa7}.

\section{The Radiation Monitor}
\label{lobo-sec4}

LPF will be stationed for operations in a Lissajous orbit around the
Earth-Sun Lagrange point L1. This is about 1.5~million kilometres
from Earth, i.e., well beyond its radiation belts. The spacecraft
will thus be exposed to particles in the solar wind, but also to
others, mostly of galactic origin. The composition of these fluxes
in charged particles is roughly 90\% protons, 8\% Helium ions, and
2\% heavier nuclei (Carbon and higher $Z\/$) and electrons. These
will charge the spacecraft through both direct and indirect deposition,
since \emph{secondary} particles are generated as the \emph{primary}
ones travel across the various materials. What creates concerns about
charging is actually only the TMs, as charge depositions in them
generates spurious potentials, hence noise in the GRS readout due
to the random character of those depositions.

\begin{figure}[t!]
\centering
\includegraphics[width=0.48\columnwidth]{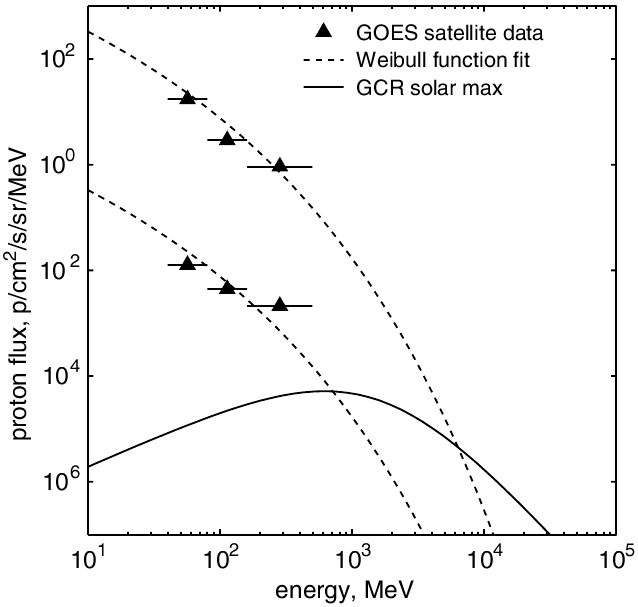}\quad
\includegraphics[width=0.48\columnwidth]{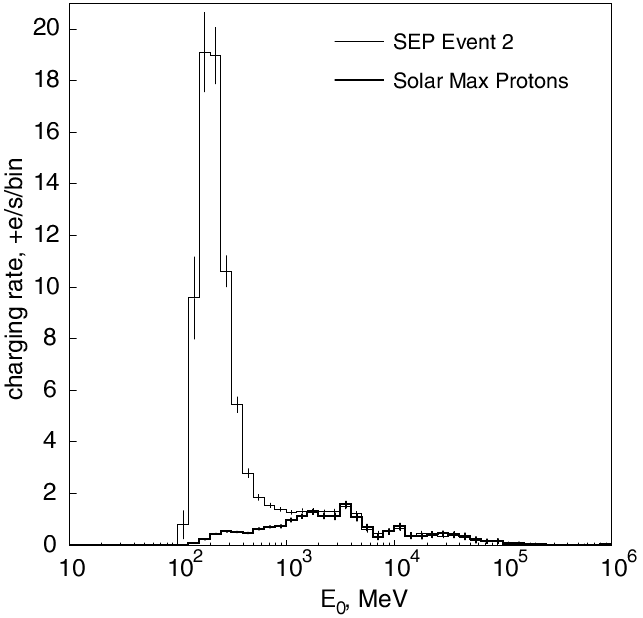}
\caption{Left panel: spectra of primary energies in a giant SEP
event (September 1989) and of a normal one (May 2001) as obtained
from the GOES Observatory data, and fitted. Also shown is a GCR
proton spectrum taken at solar maximum. Right panel: charging
rates in the TMs for a ``normal'' SEP event and GCR, also at
solar maximum.
\label{lobo-fig3}}
\end{figure}

Extensive simulations done at Imperial College~\cite{lobo-lpfsimul} with
the public CERN tool \textsl{Geant-4} show that only primary particles
with energies roughly above 70\,MeV can possibly make it to the TM; lower
energy ones are deterred by the satellite structures surrounding the
TMs, and hence can cause no harm to the experiment. The simulations
also show that the charging \emph{rates} of the TMs depends on whether
the primary particles are Galactic Cosmic Rays (GCR) or correspond to
flares in the Sun (SEP events, Solar Energetic Particles\footnote{
The solar wind contains mostly keV particles, only a relatively
small fraction are above 100~MeV, hence the name SEP.}).
These two types of fluxes have different energy spectra, and the way
to distinguish between them is therefore to do spectroscopy.
Figure~\ref{lobo-fig3} shows quite clearly the above distinction: spectra
of GCR are displayed on the left and deposition rates in the right panel.
Data are due to Ara\'ujo \emph{et al}~\cite{lobo-ezelviro}. See also
Catia Grimani's contribution to this volume,

\begin{figure}[t!]
\centering
\includegraphics[width=0.49\columnwidth]{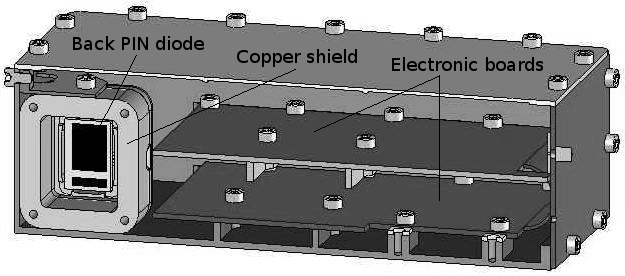}\quad
\includegraphics[width=0.47\columnwidth]{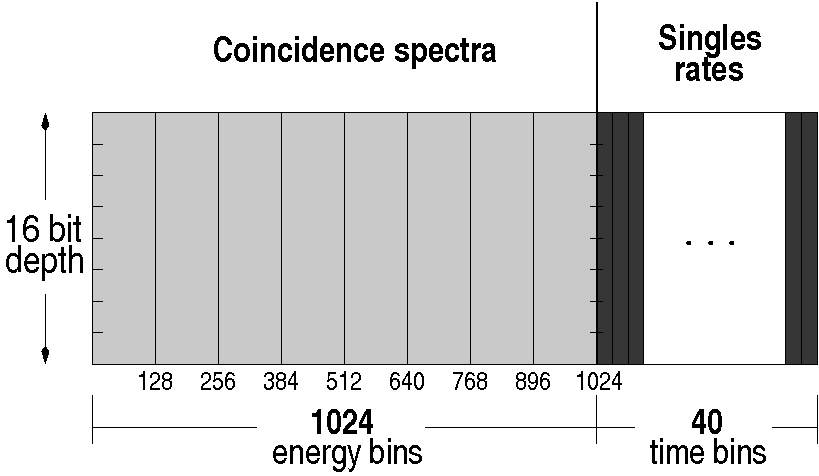}
\caption{Left panel: schematics of the Radiation Monitor. Right panel:
the Radiation Monitor data histogram, including 1024~energy bins for
coincident events and 40~bins for singles counts ---see text for
more details.
\label{lobo-fig4}}
\end{figure}

The Radiation Monitor's role in LPF is to identify the charging
rates in the TMs by short term monitoring of the charged incoming
particles. For this, a sensor consisting in two PIN diodes in
telescopic configuration are used. They are enclosed in a copper
shielding which prevents primaries with energies above 70~MeV to
reach them, thereby recreating the actual situation in the TMs,
see Figure~\ref{lobo-fig4}, left panel. The PINs are attached to
electronic circuits capable of counting both single events in either
diode as well as coincident events in the two of them. These coincidences
are used to do spectroscopy based on the energy deposition, which is
also measured. The RM data are stored in a histogram format as
shown in Figure~\ref{lobo-fig4}, right panel.

One such histogram is the RM telemetry unit. Bins are scanned and
filled at 100~Hz. Data are accumulated in the DMU during 600~seconds
(i.e., 10~minutes) and then sent to the On Board Computer (OBC) to
be telemetered to Earth according to mission schedules. Data are
subsequently analysed off-line to look for correlations with the
Charge Management System, which takes much longer periods to
produce data. We can thus use the RM to detect shorter term
charging fluctuations as well as modulations which should be
useful to more thoroughly assess the whole process~\cite{lobo-tim}.

At the time of writing, the LPF RM Flight Model is a few weeks from
final delivery. Many tests have been done on it before that: extensive
electronics tests, magnetic moment measurements, vibration, temperature
stability of the RM performance and proton irradiation. The latter was
carried out at the Paul Scherrer Institute (Switzerland) where a proton
beam facility was used to generate low dose fluxes and assess its
response for various energies and beam incidence angles. The prototype
was also submitted to such test, except it was much more aggressive
at that time, as many more things had to be checked then. Now we are
reassured everything is in place and the test can be milder to avoid
deterioration of the device before flying. Analysis of the irradiation
results will soon be complete.

\section{The DMU}
\label{lobo-sec5}

\begin{figure}[t!]
\centering
\includegraphics[width=0.79\columnwidth]{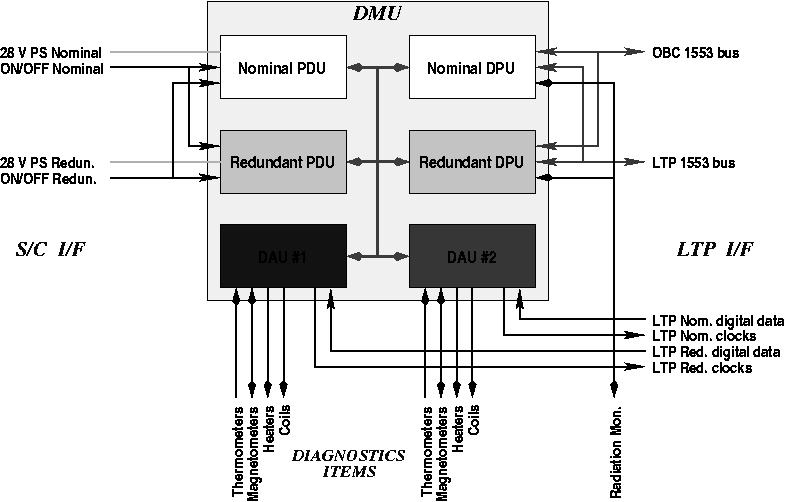}
\caption{Schematics of the DMU boards and their LTP interfaces.
\label{lobo-fig5}}
\end{figure}

The Data Management Unit (DMU) is the LTP computer, and has been
designed, developed, manufactured and delivered by IEEC, Barcelona.
The latter happened on 11~February 2010. The DMU has three electronic boards:
the Power Distribution Unit (PDU), the Data Acquisition Unit (DAU)
and the Data Processing Unit (DPU), each of them duplicated for
redundancy security. Redundancy is however not full in the DAUs,
where some diagnostics items are connected to only one DAU, see
Figure~\ref{lobo-fig5}, again due to mass-energy quotas. Protection
against possible DAU failure has been maximised by careful distribution
of each DAU's tasks.

The DMU fully controls of all the diagnostics items functionalities,
including activation, acquisition, and telemetry as described in sections
\ref{lobo-sec2} and \ref{lobo-sec3} above. With the exception, however, of
the Radiation Monitor, which internally acquires its data and communicates
with the DMU via a serial RS\,422 line. But the DMU has many more interfaces
with other parts of the LTP, as shown in Figure~\ref{lobo-fig5}.
Communications with the OBC and non-diagnostics parts of the LTP is
done via two \textsl{MIL-STD-1553} buses, respectively.

The DMU Flight Model hardware has undergone extensive testing before its
formal acceptance by the Mission Architect:

\begin{itemize}
 \setlength{\itemsep}{-0.2 ex}
 \item All Diagnostics Items performance/functional tests
 \item Mechanical tests: Centre of Gravity, vibration, and pyrotechnic shock
 \item Thermal vacuum tests
 \item Electromagnetic Compatibility tests
 \item Dipole magnetic moment measurements
 \item Magnetic stability tests
\end{itemize}

\begin{figure}[h!]
\centering
\includegraphics[width=0.65\columnwidth]{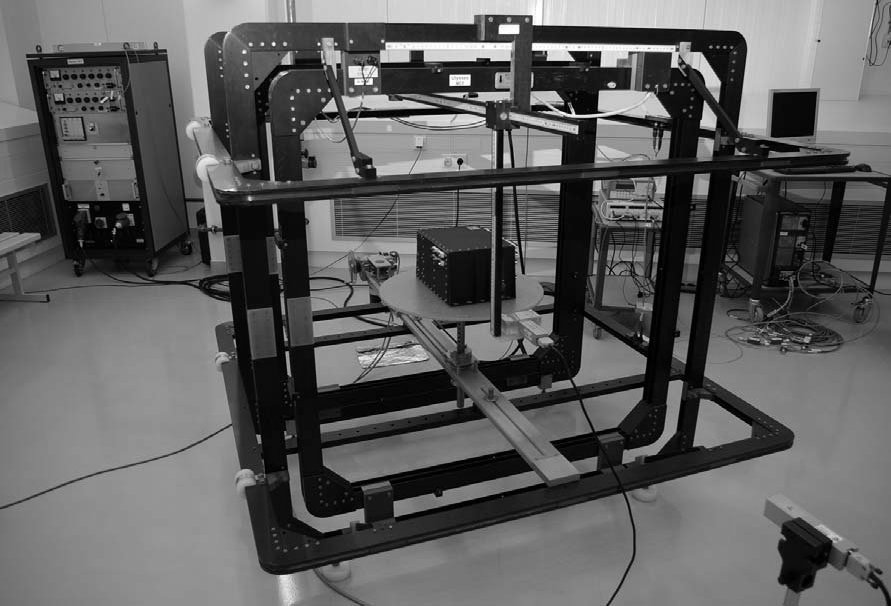}
\caption{The DMU in a test facility at ESTEC. It is the dark box at
the centre of a Helmholtz coil set, which shields the DMU from external
disturbances during magnetic moment measurements.
\label{lobo-fig6}}
\end{figure}

Some of them needed external facilities, such as magnetic tests done at
ESTEC (see Figure~\ref{lobo-fig6}), and EMC tests at a local industrial
company. All the above tests, however, have also been passed by the rest
of the DDS hardware, they are indeed not specific to the DMU.

\subsection{The DMU Software}
\label{lobo-sec5-1}

The DMU requires software to implement its functionalities and to
communicate with its various interfaces. This software has been
created from scratch at IEEC, too, and has two main bodies: the
Boot Software (BSW) and the Application Software (ASW).

\begin{figure}[t!]
\centering
\includegraphics[width=0.46\columnwidth]{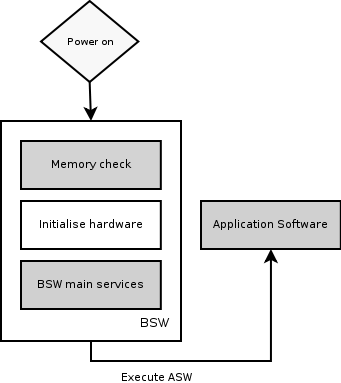}\qquad
\includegraphics[width=0.47\columnwidth]{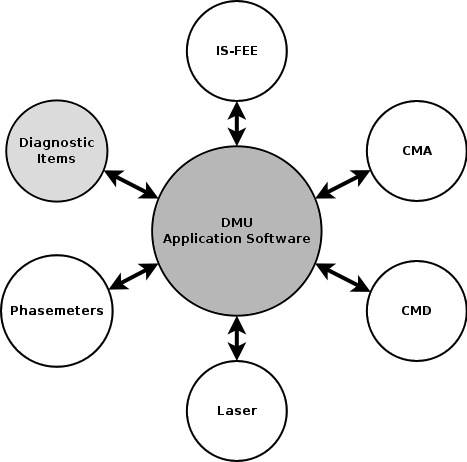}
\caption{Conceptual diagram of the DMU software: BSW on the left and
ASW on the right.
\label{lobo-fig7}}
\end{figure}

The BSW is a minimal, non-real time operating system, which occupies
61 kilobytes of memory. It contains some 30\,000~lines of $C\/$ code,
plus about 1\,000 of Assembler code, and it was eventually burnt to
a PROM, of course after extremely rigorous tests were passed, as the
PROM cannot be either rewritten or replaced. The BSW starts the DMU
and acts as scheduler, interrupt control and error detection and
correction. A most important task for the mission which is also in
charge of the BSW is to load the ASW, which understands and
executes all the telecommands which are received from the Mission
Operation Centre (MOC) following the mission Master Plan. The BSW
can also patch the ASW should there be an upgrade deemed necessary
for the continuation of the mission operations. The BSW has long
been burnt to an FM PROM, hence is fully ready for flight.

The ASW is built on top of RTEMS (Real-Time Executive for Multiprocessor
Systems), a real-time operating system, in order to cope with the hard
real-time constraints (i.e. task deadlines must be met on time) of the
mission. The ASW, as already mentioned, manages all the scientific and
technological functions of the DMU. It supports real time tasks up to
a maximum frequency of 100~Hz, and is encoded in some 70\,000~lines of
$C\/$~code, plus 300~of Assembler. In addition, it sends telemetry to
the OBC via a \textsl{MIL-STD-1553} bus. The ASW is stored in EEPROM
(Electrically Erasable Programmable Read-Only Memory), which makes
possible to receive upgrades from ground if necessary. It is executed
from RAM. Stress tests done on the ASW show that it produces a maximum
CPU load of 86\%, therefore a bit tight on margin. Work is in progress
to bring down that ratio by a few percentile points.

In addition to the mentioned stress tests, the ASW has undergone unit
testing and validation, the latter by an independent entity which
produced test scripts to check nominal test cases. Errors were filed
and reported for debugging. The latest validated version is 2.3. ASW~2.4
was released in mid August 2010, with all functionalities in it. By the
end of 2010 a validated and optimised version of ASW~2.4 will be ready.

The SW is a dynamic component of the DMU, and work on it is therefore
envisaged all the way to mission completion.

\section{Conclusions}
\label{lobo-sec6}

We have briefly reviewed here the essential parts, as well as their
functions, of the DDS. As shown, the Spanish hardware contribution to
LPF is really complete, up to minor changes/details. The IEEC team keeps
still active, and will be until the end of the mission. New tasks, not
reported here, include the preparation of the mission operations,
software (not to be confused with the DMU software) modelling of the
system response to validate the PORs (Payload Operations Requests)
which constitute the basis for the entire mission Experiment Master
Plan, and which are being prepared in the Science and Technology
Operations Centre (STOC) in ESAC, near Madrid. In addition, on-line
response and off-line data analysis of the mission yield has a lot
to do with the diagnostics subsystem, hence IEEC is very actively
involved in these matters, too.

\ack

Financial support from the Spanish Ministry of Innovation and Research
(MICINN), contract ESP2007-61712, is gratefully acknowledged, as well as
AYA08-04211-C02-01. CFS acknowledges support from the Ram\'on y Cajal
Programme of the same Ministry, and a Marie Curie International 
Reintegration Grant (MIRGCT-2007-205005/PHY) within the 7th European
Community Framework Programme, and PC acknowledges an FPU PhD Grant from
MICINN. Part of this work was also supported by AGAUR, Generalitat de
Catalunya.

\end{document}